\documentclass[]{amsart}
\usepackage{amsmath,amsfonts,amssymb}
\usepackage{verbatim}
\usepackage{enumerate}
\usepackage{url}
\def\N{\mathbb N}

\def\F{\mathbb F}
\def\ord{\mathop{\rm ord}\nolimits}
\def\rad{\mathop{\rm rad}}
\def\lcm{\mathop{\rm lcm}}

\theoremstyle{plain}
\newtheorem{theorem}{Theorem}[section]
\newtheorem{lemma}[theorem]{Lemma}

\newtheorem{corollary}[theorem]{Corollary}
\newtheorem{proposition}[theorem]{Proposition}
\newtheorem{remark}[theorem]{Remark}

\def\proof{{\it Proof: }}
\def\qed{\hfill\hbox{$\square$}}

\theoremstyle{definition}

\author[F.E. Brochero Mart\'{\i}nez]{F. E. Brochero Mart\'{\i}nez}
\author[C. R. Giraldo Vergara]{C. R. Giraldo Vergara}
\author[L. Batista de Oliveira]{L. Batista de Oliveira}
\address{
Departamento de Matem\'{a}tica\\
Universidade Federal de Minas Gerais\\
UFMG\\
Belo Horizonte, MG\\
 30123-970\\
 Brazil\\
 }
 \email{fbrocher@mat.ufmg.br }\email{carmita@mat.ufmg.br}\email{liubatista20@gmail.com}

\date{\today
}

\subjclass[2000]{}

\title{Explicit factorization of $x^n-1\in \F_q[x]$ 
}
\begin{document}

\begin{abstract}
Let $\F_q$ be a finite field  and $n$ a positive integer. In this article, we prove that, under some conditions on $q$ and $n$, the polynomial $x^n-1\in \F_q[x]$ can be split into irreducible binomials  $x^t-a$ and an explicit factorization into irreducible factors is given. 

 Finally,   weakening one of our hypothesis, we also obtain  factors of the form $x^{2t}-ax^t+b$ and explicit  splitting of   $x^n-1$  into irreducible factors.
\end{abstract}

\maketitle

\section{Introduction}
The factorization of  a polynomial over a finite field has  theoretical and practical important consequences in a wide variety of technological situations, including efficient and secure communications, error-correcting codes, deterministic simulations of random processes and digital tracking systems (see for instance  \cite{Gol}); in particular, each irreducible factor of $x^n-1$ in $\F_q[x]$ determines a cyclic code of length $n$ over $\F_q$ (see  \cite{Lin}). In fact,  each cyclic code of  length $n$ can be represented as an ideal of the ring ${\mathcal R}_n=\F_q[x]/\langle x^n-1\rangle$, and each ideal of $\mathcal R_n$ is generated by a unique factor of $x^n-1$. 

It is well known that $x^n-1=\prod_{d|n} \Phi_d(x)$, where $\Phi_d(x)$ denotes  the $d$-th cyclotomic polynomial (see \cite{LiNi} Theorem 2.45). It follows that the factorization of $x^n-1$ depends strongly on the factorization of the cyclotomic polynomial.    
In general,  a ``generic efficient algorithm''  to split   $\Phi_d(x)$ in $\F_q[x]$ for arbitrary $d$ and $q$ is an open problem and just some particular cases are known.
The complete factorization of $\Phi_{2^k}(x)=x^{2^k}+1$  over $\F_q[x]$ with $q\equiv 1 \pmod 4$ is a classical result, and in the case where $q\equiv 3 \pmod 4$ the factorization  was obtained way by Meyn in \cite{Mey}  within an elementary; in \cite{FiYu} Fitzgerald and Yucas studied how to find the explicit factorization of $\Phi_{2^kr}(x)$, where $r$ is an odd prime and $q\equiv \pm 1 \pmod r$, in particular, they obtained the explicit factorization in the case where $r=3$. 
 In   \cite{WaWa},  Wang and  Wang give a complete explicit factorization of $\Phi_{2^n\cdot 5}(x)$.  
 Finally, in \cite{CLT}  Chen, Li and Tuerhong found the explicit factorization of $x^{2^mp^n}-1$  over $\F_q$, where $p$ is an odd prime with $q\equiv 1 \pmod p$.

In this paper, we characterize every irreducible factor of $x^n-1$ over $\F_q$ assuming that  every prime divisor of $n$ is a divisor of $q-1$, and we	 give explicit expressions for the factorization of $x^n-1$,  so generalizing the results  listed above.
Observe  that, in this case,  all factors are binomials or trinomials, thus every irreducible factor is a sparse polynomial (polynomial with few nonzero term) and this type of polynomials  has important applications, for instance,  for an efficient hardware implementation of  feedback shift registers (see \cite{Ber1}).

\section{Preliminaries} 

Throughout this paper, $\F_q$ will denote a finite field of order $q$, where $q$ is a power of a 
prime,  $\theta$  is a generator of the cyclic group $\F_q^*$;  for each $a\in \F_q^*$, $\ord_q a$  will denote  the minimum positive integer $k$ such that $a^k=1$.
For each prime  $p$   and each integer $m$, $\nu_p(m)$ will  denote the maximum power of $p$ that divide $m$ and  $\rad(m)$ denotes the radical of $m$, i.e.,   if $m=p_1^{\alpha_1}p_2^{\alpha_2}\cdots p_l^{\alpha_l}$ is the factorization of $m$ in prime factors, then $\rad(m)=p_1p_2\cdots p_l$.

It is easy to see that if $n$ divides $q-1$  then $\F_q^*$ contains a primitive $n$-th root of  unity $\zeta_n=\theta^{(q-1)/n}\in \F_q^*$, and the polynomial $x^n-1$  splits into linear factor  of the form $x-\zeta_n^j$, where $j=1,2,\dots, n$, thus each factor is a binomial. 
A natural question  is when $x^n-1$ can be factorized in a way that each irreducible factor is a  binomial of the form $x^t-a$.   This question can be divide in three parts:
\begin{itemize}
\item Determine when $x^t-a$ is an irreducible  polynomial.
\item  Determine when $x^t-a$ is a factor of $x^n-1$. 
\item Find conditions over $n$ and $q$ such that every irreducible factor is a binomial.
\end{itemize}
 Observe that the first part is a classical remarkable result:

 \begin{lemma}[Theorem 3.75 in \cite{LiNi}] \label{irreducible} Let $t\ge 2$ be an integer and $a\in \F_q^*$. Then the binomial $x^t-a$ is irreducible in $\F_q[x]$ if and only if the following three conditions are satisfied: 
\begin{enumerate}
\item  $\rad(t)$ divides $\ord_q a$.
\item $\gcd(t, \frac {q-1}{\ord_q a})=1$.
\item   if $4|t$  then $q\equiv 1 \pmod 4$.
\end{enumerate}
\end{lemma}

\begin{remark}  If $t=4s$, $q\equiv 3 \pmod 4$ and $a$ is not a square in $\F_q$, then $x^t-a=(x^{2s}-bx^s+c)(x^{2s}+bx^s+c)$, where $b$ and $c$ are elements of $\F_q$ such that  $c^2=-a$ and  $b^2=2c$.
\end{remark}

For the second question, we can find elementary conditions over $t$ and $a$:
\begin{lemma}\label{divide} The binomial $x^t-a\in \F_q[x]$ is a factor of $x^n-1$ if and only if  $t|n$ and $\ord_q a|\gcd(q-1, \frac nt)$.
\end{lemma}

\proof  Using the division algorithm we know that $n=tm+r$ where $0\le r<t$. Thus
$$x^n-1=(x^t)^m x^r-1\equiv a^mx^r-1 \mod{(x^t-a)},$$
so, it follows that $x^t-a$ divides $x^n-1$ if and only if  it  also  divides  $a^mx^r-1$. Since $\deg(a^mx^r-1)<\deg(x^t-a)$, this happens if and only if  $r=0$ and $a^m=1$. 
Therefore $t$ is a divisor of $ n$ and $\ord_q a|m=\frac nt$. Since $\ord_q a|(q-1)$, we conclude that $\ord_q a|\gcd(q-1, \frac nt)$.\qed

The third part, that  will be solved in the next section, depends on some known elementary properties of the cyclotomic polynomial that we present, without proof, in the following proposition.

\begin{proposition}\label{cyclotomic}
Suppose that $m$ is a positive integer and take $p$ a prime such that $\gcd(pm,q)=1$.  Then in a finite field $\F_q$ the following properties of the cyclotomic polynomial are valid:
\begin{enumerate}[(a)]
\item $\Phi_{mp}(x)= \dfrac{\Phi_m(x^p)}{\Phi_m(x)}$ if   $p\nmid m$;

\item  $\Phi_{mp}(x)= \Phi_m(x^p) $ if $p \mid m$;

\item  $\Phi_{2m}(x) = \Phi_m(-x)$ if  $m\geq 3$ and $m$ is an odd number.
\end{enumerate}
\end{proposition}

\section{Factors of $x^n-1$}
In this section, we  prove, imposing some conditions on $n$ and $q$, that every irreducible factor of $x^n-1$ is a binomial,  and then, removing one of those conditions, we prove that every irreducible  factor is a binomial or a trinomial. 

The following proposition show that these condition are indeed necessary.
\begin{proposition}  If  every irreducible factor of $x^n-1\in \F_q[x]$ is a binomial of the form $x^t-a$, then $\rad(n)|(q-1)$ and either  $8\nmid n$ or $q\not\equiv 3 \pmod 4$.
\end{proposition}

\proof  Suppose  that $p$ is a prime such that  $p|n$. Since $\Phi_p(x)$ divides $x^p-1$ and $x^p-1$ divide $x^n-1$, it follows that every irreducible factor of $\Phi_p(x) $ is a binomial of the form $x^t-a$.
Since $x^t-a|(x^p-1)$, by Lemma \ref{divide} we have that $t|p$, so $t=1$ or $p$.  
In addition, by Theorem 2.47 (ii) in \cite{LiNi}, every irreducible factor of $\Phi_p(x)$ has degree $\ord_p q$ which is a divisor of $\varphi(p)=p-1$.    It  follows that $t=\ord_p q$ is a common divisor of $p$ and $p-1$, therefore $t=1$. It follows that  $\ord_pq=1$ or equivalently  $p|(q-1)$.

On the other hand, if $8|n$  then  $x^8-1$ divides $x^n-1$ and  if $q\equiv 3 \pmod 4$  then $-1$ is not a square in $\F_q$. 
In the case that $2$ is a square in $\F_q$, we have that
$$x^8-1=(x-1)(x+1)(x^2+1)(x^2+bx+1)(x^2-bx+1),$$
where $b^2=2$. The case that $2$ is not a square in $\F_q$, then $-2$ is a square, therefore
$$x^8-1=(x-1)(x+1)(x^2+1)(x^2+bx-1)(x^2-bx-1),$$
where $b^2=-2$. So there exist irreducible factors of $x^n-1$  which  are not  binomials. \qed

The following result shows that these conditions are in fact sufficient conditions.

\begin{theorem}\label{typefactor}
 Let $\F_q$ be a finite field  and $n$ a positive integer such that 
\begin{enumerate}
\item $q\not\equiv 3 \pmod 4$ or $8\nmid n$. 
\item $rad(n)|(q-1)$.
\end{enumerate}
Then every irreducible factor of $x^n-1$ in $\F_q[x]$ is of the form $x^t-a$, where $t\in \N$ and $a\in \F_q$ satisfy the condition of Lemmas \ref{irreducible} and \ref{divide}.
\end{theorem}

\proof We  proceed by induction over $\Omega(n)$, the total number of prime power factors of  $n$  (i.e., $\Omega(n)=\alpha_1+\cdots+\alpha_k$ where  $n=p_1^{\alpha_1}\cdots p_k^{\alpha_k}$). 

Observe that if $n|(q-1)$, then there exists an element $\zeta_n\in \F_q$, a primitive $n$-th root of  unity in $\F_q$, and  in this case 
$$x^n-1=\prod_{j=0}^{n-1} (x-\zeta_n^j),$$
so, the theorem is clearly true. In particular, if $\Omega(n)=1$ then $n=\rad(n)$ is a prime that divides $q-1$, therefore the first step of the induction is true. 

Suppose now that the theorem is true  for all $n$ such that $\rad(n)|(q-1)$ and $\Omega(n)\le N\in \N$ for some $N\ge 1$.

Let $n$ be an integer such that $rad(n)|(q-1)$ and $\Omega(n)=N+1$.  
Observe that
$$x^n-1=\prod_{d|n} \Phi_d(x)=\Phi_n(x) \cdot\prod_{{d|n \atop d\ne n}} \Phi_d(x), $$
and for all $d|n$, $d\ne n$ we have that $\Phi_d(x)|(x^d-1)$ Then, by induction hypothesis  we have that every irreducible factor of $\Phi_d(x)$ is of the form $x^t-a$,  so we only need to verify the condition of Lemma \ref{divide}. 
In fact, since  $t$ divides $\frac {d}{gcd(d,q-1)}$ then it  divides $\frac {n}{gcd(n,q-1)}$. In addition, $\ord_q a|\gcd(q-1, \frac dt)|\gcd(q-1, \frac nt)$.

Therefore, it is enough to analyse the factors of  $\Phi_n(x)$. 
Since the theorem is true when $n|(q-1)$, we can suppose without loss of generality that there exists a prime $p$ divisor of $n$, such that  $\nu_p(n)>\nu_p(q-1)\ge \nu_p(\rad(n))=1$. 
So we can assume that $n=pm$ with $\nu_p(m)\ge \nu_p(q-1)\ge 1$.

 At this point, we  consider two cases:
\begin{enumerate}
\item $p\ne2$ or $q\not\equiv 3\pmod 4$
\item $p=2$, $q\equiv 3 \pmod 4$ and $\nu_{p'}(n)\le \nu_{p'}(q-1)$ for each  odd prime factor $p'$ of $n$ .
\end{enumerate}

In the first case, by Proposition \ref{cyclotomic} we have
$$\Phi_n(x)=\Phi_{pm}(x)=\Phi_m(x^p).$$
Since $\Omega(m)<\Omega(n)$ it follows by induction hypothesis that every  irreducible factor of $\Phi_m(x)$ is of the form $x^t-a$ and so it   satisfies the condition of Lemmas \ref{irreducible} and \ref{divide}. In addition, 
$x^{pt}-a$ is a factor, not necessarily  irreducible,  of $\Phi_m(x^p)$. 

Observe that $\gcd(tp, \frac {q-1}{\ord_q a})=1$ or $p$. When 
 the greatest common divisor is $1$, we have that $\nu_p(\ord_q a)=\nu_p(q-1)\ge 1$, then $p|\ord_q a$, and therefore $\rad(pt)$ divides $\ord_q a$, so  it follows that the first two conditions of Lemma \ref{irreducible} are satisfied.   
In addition,  if $q\equiv 3 \pmod 4$ it   follows that $p\ne 2$, and by induction hypothesis  we know that  $4\nmid t$, therefore $4\nmid pt$,  consequently the third condition of Lemma \ref{irreducible} is also satisfied and then $x^{pt}-a$ is irreducible.
The condition of Lemma \ref{divide} are satisfied, because we have already proved that  $x^{pt}-a$ is a factor of $x^n-1$.

When  the greatest common divisor is $p$, we have that $\nu_p(q-1)>\nu_p(\ord_q a)$ and $p\nmid t$. Then  $\ord_q a|\frac {q-1}p$, or equivalently,  $a$ is a root of the polynomial $P(x)=x^{\frac {q-1}p}-1$. Since the roots of this polynomial are $\{\theta^{jp}|j=1,2,\dots,\frac {q-1}p\}$, so  there exists $b\in \F_q^*$ such that $a=b^p$,  and therefore
$$x^{pt}-a=x^{pt}-b^p=\prod_{j=1}^p (x^t-\zeta_p^j b),$$
where $\zeta_p\in \F_q$ is a primitive $p$-th  root of  unity.  Note that
$$\ord_q(\zeta_p^j b)=\lcm (\ord_p \zeta_p^j,\ord_p b)=\lcm (p,p\cdot\ord_p a)=p\cdot \ord_q a,$$
so $\rad(t)$ divides $ \ord_q(\zeta_p^j b)$ and $\gcd\left( t, \frac {q-1}{\ord_q (\zeta_p^j b)}\right)=1$.  Hence  every factor of the form $x^t-\zeta_p^j b$
is an irreducible factor of $x^n-1$, which completes the first case.

Suppose now that $p=2$, $q\equiv 3 \pmod 4$ and that for each  odd prime factor $p'$ of $n$ we have that $\nu_{p'}(n)\le \nu_{p'}(q-1)$.  By the hypothesis,  we know that $$3> \nu_2(n)>\nu_2(q-1)=1,$$
 so $n=4m$ where $m$ is an odd number such that $m|(q-1)$. In the case $m=1$, the cyclotomic polynomial  $\Phi_4(x)=x^2+1$ is irreducible, because $-1$ is not a square in $\F_q$. 
Thereby, we can suppose that $m\ge 3$ and  by Proposition \ref{cyclotomic} we have
$$\Phi_{n}(x)=\Phi_{2m}(x^2)= \Phi_m(-x^2).$$
Now, since $\Omega(m)< \Omega(n)$, by induction hypothesis we know that every irreducible factor of $\Phi_m(x)$ is of the form $x^t-a$,  so it  satisfies  the condition of Lemmas  \ref{irreducible} and \ref{divide},  and   $x^{2t}-(-a)$ is a factor of $\Phi_n(x)$. We claim that it is an irreducible factor.   
In fact, since 
$$\ord_q(-a)=\lcm( \ord_q(-1), \ord_q(a)),$$ then $\ord_q(-a)$ is an even number  and then $\rad(2t)$ divides $\ord_q(-a)$. In addition, $\frac {q-1}{\ord_q(-a)}$ is an odd number, therefore $\gcd\left(2t, \frac {q-1}{\ord_q(-a)}\right)=1$.  Finally, since $t|m$ then $2t$ is not divisible by $4$, thus by Lemma \ref{irreducible}  it follows that $x^{2t}-(-a)$ is an irreducible factor of $x^n-1$. \qed

\begin{corollary}\label{explicit1} Let $n$ and $q$ be as in  Theorem \ref{typefactor} and set  $m=\frac n{\gcd(n, q-1)}$ and $l=\frac {q-1}{\gcd(q-1,n)}$. Then
\begin{enumerate}[(a)]
\item 
The factorization of $x^n-1$ into irreducible factors of $\F_q[x]$ is
$$\prod_{t|m}\prod_{{1\le u\le \gcd(n,q-1)\atop \gcd(u,t)=1}} (x^t-\theta^{ul}).$$
\item For each $t|m$, the number of irreducible factors of degree $t$ is $\frac{\varphi(t)}t\cdot \gcd(n,q-1)$,  where $\varphi$ denotes the Euler Totient function and the total number of irreducible factors is 
$$ \gcd(n,q-1)\cdot\prod_{p|m\atop p{\text{ prime}}} \left(1+\nu_p(m)\frac {p-1}p\right).$$
\end{enumerate}
\end{corollary}

\proof  {\em (a)} \quad Let $x^t-a$ be an arbitrary  irreducible factor of $x^n-1$, 
and let $p$ be a prime such that $p|t$.  As a consequence of Lemma \ref{irreducible}  we have that 
$$\nu_p(\ord_q a)\ge 1\text{ and } \nu_p(q-1)=\nu_p(\ord_q a).$$
In addition,  as a consequence of Lemma \ref{divide} we have that
$$\nu_p(t)\le \nu_p(n) \text{ and } \nu_p(\ord_q a)\le \min\left\{ \nu_p(q-1), \nu_p\left(\frac nt\right)\right\}\le \nu_p(n)-\nu_p(t).$$
Therefore,  for each prime $p$ such that $ \nu_p(t)\ge 1$, we have that    $\nu_p(t)\le\nu_p(n)-\nu_p(q-1),$ 
so  we conclude   that $t$ divides  $m=\frac {n}{\gcd(n, q-1)}$.

Now, note that   $a=\theta^v$  for some  $0\le v\le q-1$,  and since
$$\nu_p(q-1)=\nu_p(\ord_q a)=\nu_p(\ord_q \theta^v)=\nu_p\left( \frac {q-1}{\gcd(q-1,v)}\right)=\max\{ 0, \nu_p(q-1)-\nu_p(v)\},$$
we have that $\nu_p(v)=0$, so  it follows that $\gcd(t, v)=1$.

Using  that $\ord_q a|\frac nt$, we have that  $a^{n/t}=\theta^{vn/t}=1$, then $vn/t$ is a multiple of the order of $\theta$, i.e.  $(q-1)|(vn/t)$, and therefore $\left.\frac {q-1}{\gcd(q-1,n/t)}\right|v$.

Since $t$ is a divisor of $\frac {n}{\gcd(n, q-1)}$, we have that $\frac {q-1}{\gcd(q-1,n/t)}$ is a divisor of $l=\frac {q-1}{\gcd(q-1,n)}$. On the other hand, note that   $\frac {q-1}{\gcd(q-1,n/t)}$ is  a multiple of
$$\frac {q-1}{\gcd(q-1,n/m)}=\frac {q-1}{\gcd(q-1,\gcd(n,q-1))}=l.$$
therefore 
$$\frac {q-1}{\gcd(q-1,n/t)}=\frac {q-1}{\gcd(q-1,n)}.$$
It follows that $l$ divides $v$ whatever is  $t$, thus we can set $v=ul$, where $1\le u\le \frac{q-1}l=\gcd (n,q-1)$.

Finally, since $\gcd(t,u)|\gcd(t,v)=1$, we conclude that $\gcd(t,u)=1$.  
This facts show that $x^n-1$ divide $\prod\limits_{t|m}\prod\limits_{{1\le u\le \gcd(n,q-1)\atop \gcd(u,t)=1}} (x^t-\theta^{ul}).$

Conversely, if $t$ and $u$  satisfy these conditions on the product, it can verified directly that $x^t-\theta^{ul} $ is an irreducible factor of $x^n-1$, so {\em (a)} is proved.

{\em (b)} Let $t$ be a divisor of $m=\frac n{\gcd(n,q-1)}$. Since
 $$\rad(t)|\rad(n)|\gcd(q-1,n),$$
 every prime that divides $t$ also divides $\gcd(q-1,n)$.
Let   $p_1,p_2,\dots, p_k$ be   the list of primes that divide $t$. It follows that  there exist $(1-\frac 1{p_1}) \cdot \gcd(q-1,n)$ numbers less or equal to $\gcd(q-1,n)$ that do not have any common  factor  with $p_1$, that there exist  $(1-\frac 1{p_1}) (1-\frac 1{p_2}) \cdot \gcd(q-1,n)$
 that do not have any common  factor with $p_1$ and $p_2$, and inductively we conclude that there exist
$$\left(1-\frac 1{p_1}\right)\left (1-\frac 1{p_2}\right)\cdots \left(1-\frac 1{p_k}\right) \cdot \gcd(q-1,n)=\frac {\varphi(t)}{t} \gcd(q-1,n)$$ 
numbers without any common factor  with $t$.
Finally the total number of irreducible factors of $x^n-1$  is
$$\sum_{t|m} \frac {\varphi(t)}{t} \cdot  \gcd(q-1,n).$$
Observe now that   the function $\frac{\varphi(t)}t$ is a multiplicative function, therefore $\sum_{t|m} \frac {\varphi(t)}{t}$ is also multiplicative  and thus it is enough to calculate this sum for powers of primes.  In this case 
we have that
$$\sum_{d|p^k} \frac {\varphi(p^k)}{p^k}=1+k\left(1-\frac 1p\right),$$
obtaining the part {\em (b)}.
 \qed

In the following theorem,  we analyze the factorization of $x^n-1$  in the case when $8|n$ and $q\equiv 3 \pmod 4$.

\begin{theorem}\label{case4k+3}
 Let $\F_q$ be a finite field  and $n$ a positive integer such that  
\begin{enumerate}
\item $q\equiv 3 \pmod 4$ and $8\mid n$. 
\item $rad(n)|(q-1)$.
\end{enumerate}
Then each irreducible factor of $x^n-1$ in $\F_q[x]$ is of one  of the following types:
\begin{enumerate}[(a)]
\item $x^t-a$ satisfying  the hypothesis of Lemmas  \ref{irreducible} and \ref{divide}. 
\item $x^{2t}+(a+a^q)x^t+a^{q+1}\in \F_q[x]$, where $a\in \F_{q^2}\setminus \F_q$ and $x^t-a\in \F_{q^2}[x]$  satisfies the hypothesis of Lemmas  \ref{irreducible} and \ref{divide} in $\F_{q^2}$.
\end{enumerate}
\end{theorem}

\proof
Since $q^2 \equiv 1 \pmod 4$, it follows that the irreducible factors of $x^n-1$ in $\F_{q^2}[x]$  are of the form $x^t-a$ and then they  satisfy    Lemmas  \ref{irreducible} and \ref{divide} in $\F_{q^2}[x]$. 
Now, let $f(x)\in \F_q[x]$ be a monic irreducible  factor of $x^n-1$. From the fact that $f(x)$ is also a polynomial in $\F_{q^2}[x]$, we have two cases to consider:
\begin{enumerate}[{\em i)}]
\item If  $f(x)$ is irreducible in $\F_{q^2}[x]$, then $f(x)=x^t-a$, where $a\in \F_q$, thus it satisfies the conditions of Lemmas   \ref{irreducible} and \ref{divide} in $\F_{q}$.
\item If  $f(x)$ is not irreducible in $\F_{q^2}[x]$, then there exists $a\in \F_{q^2}\setminus \F_q$  such that $(x^t-a)|f(x)$ . Since $\sigma_q(f(x))=f(x)$, where $\sigma_q$ is  the Frobenius automorphism 
$$\begin{array}{cccc}
\sigma_q:&\F_{q^2}&\to&\F_{q^2}\\
&b&\mapsto&b^q
\end{array},$$
 it follows that $\sigma_q(x^t-a)=x^t-a^q$ also divides $f(x)$. But $a\ne a^q$, therefore $(x^t-a)(x^t-a^q)|f(x)$.
 Finally, note that $(x^t-a)(x^t-a^q)$ is invariant by $\sigma_q$, hence $(x^t-a)(x^t-a^q)\in \F_q[x]$, and since $f(x)$ is irreducible,
we conclude that $f(x)=x^{2t}-(a+a^q)x^t+a^{q+1}$.\qed
\end{enumerate}

\begin{remark} With an equivalent  proof of theorem above, but replacing the the hypotheses of the theorem by   $\rad(n)|(q^2-1)$, or equivalently,  every prime factor $p$ of $n$ satisfies that $q\equiv \pm 1 \pmod p$, it is possible to prove that every irreducible factor of $x^n-1$ is a binomial or trinomial.
\end{remark}

\begin{corollary} \label{factorq^2-1} Let $n$ and $q$ be as in  Theorem \ref{case4k+3}. Let $\alpha$ be a generator of the cyclic group $\F_{q^2} ^*$ 
satisfying $\alpha^{q+1}=\theta$ and put $m=\frac n{\gcd(n, q^2-1)}$, $l_1=\frac {q-1}{\gcd(q-1,n)}$, $l_2=\frac {q^2-1}{\gcd(q^2-1,n)}$ and 
$r=\min\{\nu_2(n/2),\nu_2(q+1)\}$. Then
\begin{enumerate}[(a)]
\item 
The factorization of $x^n-1$ into irreducible factors of $\F_q[x]$ is
$$\prod_{t|m\atop t\text{ odd}}\prod_{{1\le w\le\gcd(n,q-1)}\atop \gcd(w,t)=1} (x^t-\theta^{wl_1})\cdot \prod_{t|m}\prod_{ u\in\mathcal R_t} (x^{2t}-(\alpha^{ul_2}+\alpha^{qul_2})x^t+\theta^{ul_2}),$$
where $\mathcal R_t$ is the set 
 $$\left\{u\in\N\left|  {1\le u\le\gcd(n, q^2-1), \gcd(u,t)=1\atop 
2^r\nmid u \text{ and }\ u< \{qu\}_{\gcd(n,q^2-1)}}\right.\right\} $$
and $\{a\}_b$ denotes the  remainder of the division of $a$ by $b$, i.e.  it is  the number $0\le c<b$ such that $a\equiv c\pmod b$.

\item For each $t$ odd with $t|m$, the number of irreducible binomials  of degree $t$ and   $2t$ are
$\dfrac{\varphi(t)}t\cdot \gcd(n,q-1)$ and
$\dfrac{\varphi(t)}{2t}\cdot \gcd(n,q-1)$    respectively,  and
the number irreducible trinomials  of degree $2t$ is
$$
\begin{cases}
\dfrac{\varphi(t)}t\cdot 2^{r-1} \gcd(n,q-1)&\text{if $t$ is even}\\
\dfrac{\varphi(t)}{t}\cdot (2^{r-1}-1) \gcd(n,q-1)&\text{if $t$ is odd}.
\end{cases}
$$
The total number of irreducible factors is 
$$ \gcd(n,q-1)\cdot \left(\frac 12+2^{r-2}(2+\nu_2(m ))\right)\cdot\prod_{p|m\atop p{\text{ odd prime}}} \left(1+\nu_p(m)\frac {p-1}p\right).$$
\end{enumerate}
\end{corollary}

\proof  Since $\rad(n)|(q-1)$, we have that the only prime factor in common between $n$ and $q+1$ is $2$, so $\gcd(n/2, q+1)=2^r$.  In addition,
$$l_2= \frac {q^2-1}{\gcd(q^2-1,n)}=\frac {q+1}{\gcd(q+1,n/2)}\frac {q-1}{\gcd(q-1,n)}=\frac {q+1}{2^r}{l_1}.$$

Now, we know by Corollary \ref{explicit1} that $x^n-1$ can be split in $\F_{q^2}[x]$ as
$$\prod_{t|m}\prod_{{1\le u\le \gcd(n,q^2-1)\atop \gcd(u,t)=1}} (x^t-\alpha^{ul_2}).$$
Note that  a factor $x^t-\alpha^{ul_2}$ is in $\F_q[x]$ if and only if $\alpha^{ul_2}$ is invarant by the Frobenius automorphism, i.e. $\alpha^{ul_2}=\alpha^{ul_2q}$. This last  equation is equivalent to saying that 
$(q^2-1)|(q-1) ul_2$, so $(q+1)| ul_2$.
On the other hand
\begin{align*}
\gcd(q+1,l_2)&=\gcd\left(q+1, \frac{q+1}{\gcd(n/2,q+1)}\cdot \frac{q-1}{\gcd(n,q-1)}\right)\\
&= \gcd\left(q+1, \frac{q+1}{\gcd(n/2,q+1)}\right)\\
&= \frac{q+1}{\gcd(n/2,q+1)}=\frac{q+1}{2^r},
\end{align*}
so $x^t-\alpha^{ul_2}$ is in $\F_q[x]$ if and only if $2^r|u$. Thus, we can write $u=2^rw$, and since $\gcd(u,t)=1$, it follows that $t$ is odd  and $\gcd(w,t)=1$.  Moreover,  from the fact that 
$$ul_2=2^rw \frac {q+1}{2^r}l_1=(q+1)wl_1,$$
we have that $x^t-\alpha^{ul_2}=x^t-\alpha^{(q+1)wl_1}=x^t-\theta^{wl_1}$, and these conditions determine every  irreducible  factor of $x^n-1$  that is binomial in $\F_q[x]$.  

For the values of $u$ such that $2^r\nmid u$, we have that $x^t-\alpha^{ul_2} \notin \F_q[x]$ and then
$$(x^t-\alpha^{ul_2})(x^t-\alpha^{qul_2})=x^{2t}-(\alpha^{ul_2}+\alpha^{qul_2})x^t+\theta^{ul_2}$$
is an irreducible trinomial of $\F_q[x]$.
Thus, each pair $(ul_2,\{uql_2\}_{q^2-1})$, or equivalently
each pair $(u,\{qu\}_{\gcd(n, q^2-1)})$, generates an irreducible trinomial. It follows that each irreducible trinomial is generated by a unique pair in the set
$$ \left\{(u,v)\left|  {1\le u\le\gcd(n, q^2-1), \gcd(u,t)=1\atop 
v=\{qu\}_{\gcd(n,q^2-1)},  2^r\nmid u \text{ and }\ u< v}\right.\right\} $$
and this concludes the part {\em (a)}.

An important fact to emphasize is that, if  $\alpha^{ul_2}+\alpha^{qul_2}=0$, then  the  ``trinomial'' $x^{2t}-(\alpha^{ul_2}+\alpha^{qul_2})x^t+\theta^{ul_2}$ is really a binomial of degree $2t$ with $t$ odd. 

Now, by Corollary \ref{explicit1}, we know that for every  $t$ divisor of $m$, the number of  irreducible binomials in $\F_{q^2}[x]$ is
$$\frac {\varphi(t)}{t} \gcd(n,q^2-1)=\frac {\varphi(t)}{t}2^r \gcd(n,q-1).$$
If $t$ is even, then $u$ is odd and there are not   binomials of degree $t$   in $\F_{q^2}[x]$ that are also in $\F_q[x]$. Thereby, for every binomial in $\F_{q^2}[x]$,  there exists  a unique  binomials in $\F_{q^2}[x]$, such that its product generates an irreducible trinomial in $\F_q[x]$. So the number of irreducible trinomials of degree $2t$  is   $\frac {\varphi(t)}{t}2^{r-1} \gcd(n,q-1)$.

Now, if $t$ is odd there exist $\frac {\varphi(t)}{t} \gcd(n,q-1)$ irreducible binomials in $\F_{q^2}[x]$ than are also in $\F_q[x]$, and
 $\frac {\varphi(2t)}{2t} \gcd(n,q-1)$ reducible binomials in  $\F_{q^2}[x]$ that are irreducible binomial in $\F_q[x]$. 
 Therefore,  there are 
$$\frac {\varphi(t)}{t} \gcd(n,q^2-1)-2\frac {\varphi(t)}{t} \gcd(n,q-1)=\frac {\varphi(t)}{t} (2^r-2)\gcd(n,q-1)$$ binomials in $\F_{q^2}[x]$ than  are not  in $\F_q[x]$,  so the number of irreducible trinomials of degree $2t$ is 
$\frac {\varphi(t)}{2t} (2^r-2)\gcd(n,q-1).$

Finally, the total number of  irreducible factors is
\begin{align*}&=\sum_{t|m\atop t\text{ odd}} \frac {\varphi(t)}{2t}(2^r+1) \cdot  \gcd(q-1,n)+\sum_{t|m\atop t\text{ even}} \frac {\varphi(t)}{t}2^{r-1}\cdot  \gcd(q-1,n) \\
&=\sum_{t|m\atop t\text{ odd}} \frac {\varphi(t)}{2t}\cdot  \gcd(q-1,n)+\sum_{t|m\atop t}\frac {\varphi(t)}{t}2^{r-1}\cdot  \gcd(q-1,n) \\
&= \gcd(n,q-1)\left(\frac 12\cdot\prod_{p|m\atop p{\text{ odd prime}}} \left(1+\nu_p(m)\frac {p-1}p\right)+ 2^{r-1}\cdot\prod_{p|m\atop p{\text{ prime}}} \left(1+\nu_p(m)\frac {p-1}p\right)\right)\\
&= \gcd(n,q-1)\cdot \left(\frac 12+2^{r-2}(2+\nu_2(m ))\right)\cdot\prod_{p|m\atop p{\text{ odd prime}}} \left(1+\nu_p(m)\frac {p-1}p\right),
\end{align*}
as we wanted to prove.\qed

{\em Acknowledgements.}  We would like to thank  anonymous  referees for the comments about  the first version of this article, in particular, by propose  the question about the converse of Theorem \ref{typefactor}.

\end{document}